\documentclass[aip, jcp,amsmath,amssymb,reprint,author-numerical,twocolumn]{revtex4}
\usepackage{epsfig}
\usepackage{graphicx}
\usepackage{dcolumn}
\usepackage{color}

\begin{document}
\title{Magneto-structural  transformations via a solid-state nudged elastic band method: Application to iron under pressure}
\author{N.~A. Zarkevich,$^{1}$  and D.~D. Johnson$^{1,2}$}
\email{zarkev@ameslab.gov;   ddj@ameslab.gov}
\affiliation{$^{1}$The Ames Laboratory, U.S. Department of Energy, Ames, Iowa 50011-3020 USA;}
\affiliation{$^{2}$Materials Science \& Engineering, Iowa State University, Ames, Iowa 50011-2300 USA.}

\date{\today}
\begin{abstract}
We extend the solid-state nudged elastic band method to handle 
a non-conserved order parameter -- in particular, magnetization,  that couples to volume and leads to many observed effects in magnetic systems.
We apply this formalism to the well-studied magneto-volume collapse during the pressure-induced transformation in iron -- from ferromagnetic body-centered cubic (bcc) austenite to hexagonal close-packed (hcp) martensite. 
We find a bcc-hcp equilibrium coexistence pressure of 8.4 GPa, with the transition-state enthalpy of 156 meV/Fe at this pressure. 
A discontinuity in magnetization and coherent stress occurs at the transition state, which has a form of a cusp on the potential-energy surface  (yet all the atomic and cell degrees of freedom are continuous); the calculated pressure jump of 25 GPa is related to the observed 25 GPa spread in measured coexistence pressures arising from martensitic and coherency stresses in samples. 
Our results agree with experiments, but necessarily differ from those arising from drag and restricted parametrization methods having improperly constrained or uncontrolled degrees of freedom. 
\end{abstract}
\maketitle

\section{\label{Introduction}Introduction}
{\par }
Magneto-structural transformations are quite common in magnetic systems  \cite{Vonsovsky1971}.
Their generic feature is a rapid change of magnetization and density, referred to as a magneto-volume collapse \cite{Akhiezer1981}.
While the nudged elastic band (NEB) \cite{NEB1998} and the solid-state nudged elastic band (SSNEB) \cite{GSSNEB} methods correctly account for all atomic degrees of freedom (DoF), they expect the transition state (TS) region in the form of a saddle. 
The oft-used drag methods can miss the correct TS (due to a possible discontinuity in some of the DoF), and a restricted parametrization can overlook it (due to a constrained search space). 
Here we generalize the TS search algorithm within the SSNEB method to make it suitable for the TS of various shapes (not only saddles, but also cusps), and apply it to iron under pressure, which displays a magneto-volume collapse of 11\% at the TS due to the loss of magnetization \cite{Vonsovsky1971},  a non-conserved order parameter that does not have to be continuous, unlike the atomic or cellular DoF.

{\par }
Iron is the most stable element produced by nuclear reactions, the most abundant element in the Earth core, and is also very common in extraterrestrial meteorites \cite{abundance1961}.
Metallic iron is known for its magnetism and technological impact. 
Deeper understanding of its properties under pressure affects metallurgy, materials science and engineering, geophysics, and planetary sciences  \cite{PRB_ironP}.

{\par }
Due to its importance, the iron bcc-hcp ($\alpha$-$\varepsilon$) transition and its mechanism have long been studied  \cite{Science238n4828y1987,JChemPhys128n10p104703y2008,PRB79n13p134113,JAP69n8p6126y1991,HyperfineInteractions72n4p375y1992,PRB58p5296y1998,PRB57p5647y1998,MultiscaleModMat538p523y1999,PRB77n17p174117y2008,MatTrans42n8p1571y2001,ActaPhysicaSinica59n7p4888y2010,JPhysCM21n49p495702y2009,JPhysCM22n43p435404y2010,AbstractsACS233p118y2007,PRL95n7p075501y2005,JPhysCM17n11pS957y2005,PRL93n11p115501y2004,HighPRes22n2p451y2002,Science267n5206p1972y1995,PropEPMHighPT101p159y1998,PRB87n2p024103y2013,PRB_ironP}.
In spite of iron's structural simplicity under pressure, the minimum-energy pathway (MEP) and the TS remained unresolved, in particular, because of the improper handling of magnetic DoF and expectation of saddle point behavior near the TS. 
We use the bcc-hcp transition in iron  as a prototype for the quantitative description of magnetic systems via the SSNEB method \cite{GSSNEB}
and its modifications \cite{C2NEB}.
For a magneto-structural transformation, we show that the TS is a cusp on the potential energy surface (PES) and is 
not a single electronic state, but a duality of a non-magnetic and magnetic states with the same atomic and cell coordinates, the same enthalpy, but different electronic structure. 
As a result of the coherency stress between those two electronic states, 
a pressure discontinuity at the TS exists, which can be related to the experimental scatter in the coexistence pressures of magnetic and non-magnetic phases.

\begin{figure}[b]
\vspace{-4mm}
\begin{center}
\includegraphics[width=70mm]{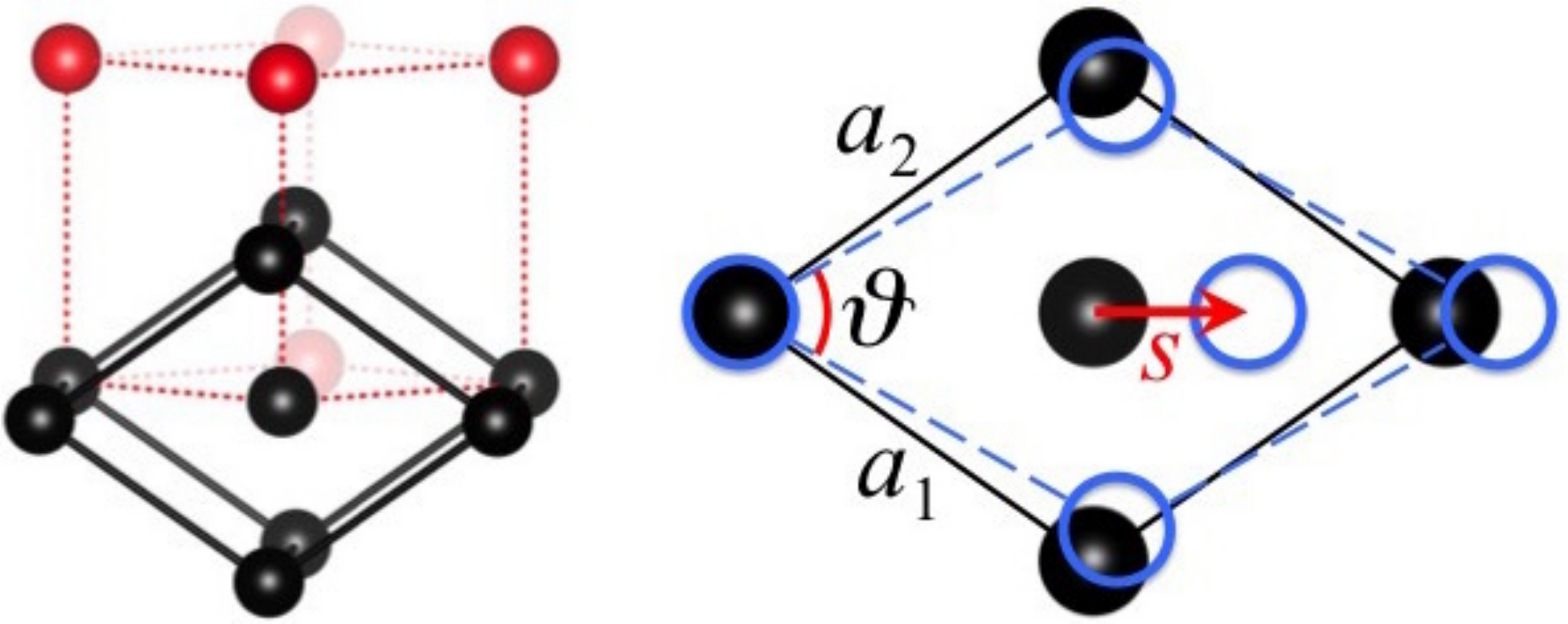}    
\end{center}
\vspace{-6mm}
\caption{\label{Fig1}(color online) 
Left: Smallest (2-atom) bcc unit cell that permits a bcc$\to$hcp transform using a shuffle-shear concept. Right: bcc$\to$hcp via shuffle-shear; bcc  ([110] projection) is shown by black atoms and solid lines, and hcp ([0001] projection) is shown as blue open circles and dashed lines, with ABAB stacking of corner A and central B atoms. Vector $\bf{s}$ (red) gives the shuffle direction from bcc to hcp. The shear angle $\vartheta$ changes from $70^\circ $ in bcc to $60^\circ $ in hcp.
}
\end{figure}

{\par }As discussed below, the simplest bcc-hcp transition in iron can be described in a 2-atom cell by one atomic and three cell (4 total) degrees of freedom, 
with the rest being constant during the whole transformation due to symmetry (Fig.~\ref{Fig1}). 
The atomic DoF in a 2-atom cell can be completely described by a continuous shuffle $s$ (which linearly changes from 0 for bcc to 1 for hcp); 
with direct coordinates of atom 1 fixed at $(0; 0; 0)$ and atom 2 at ($\frac{1}{2} [1+\frac{s}{3}$]; $\frac{1}{2} [1+\frac{s}{3}$]; $\frac{1}{2}$).

\begin{figure}[t]
\includegraphics[width=60mm]{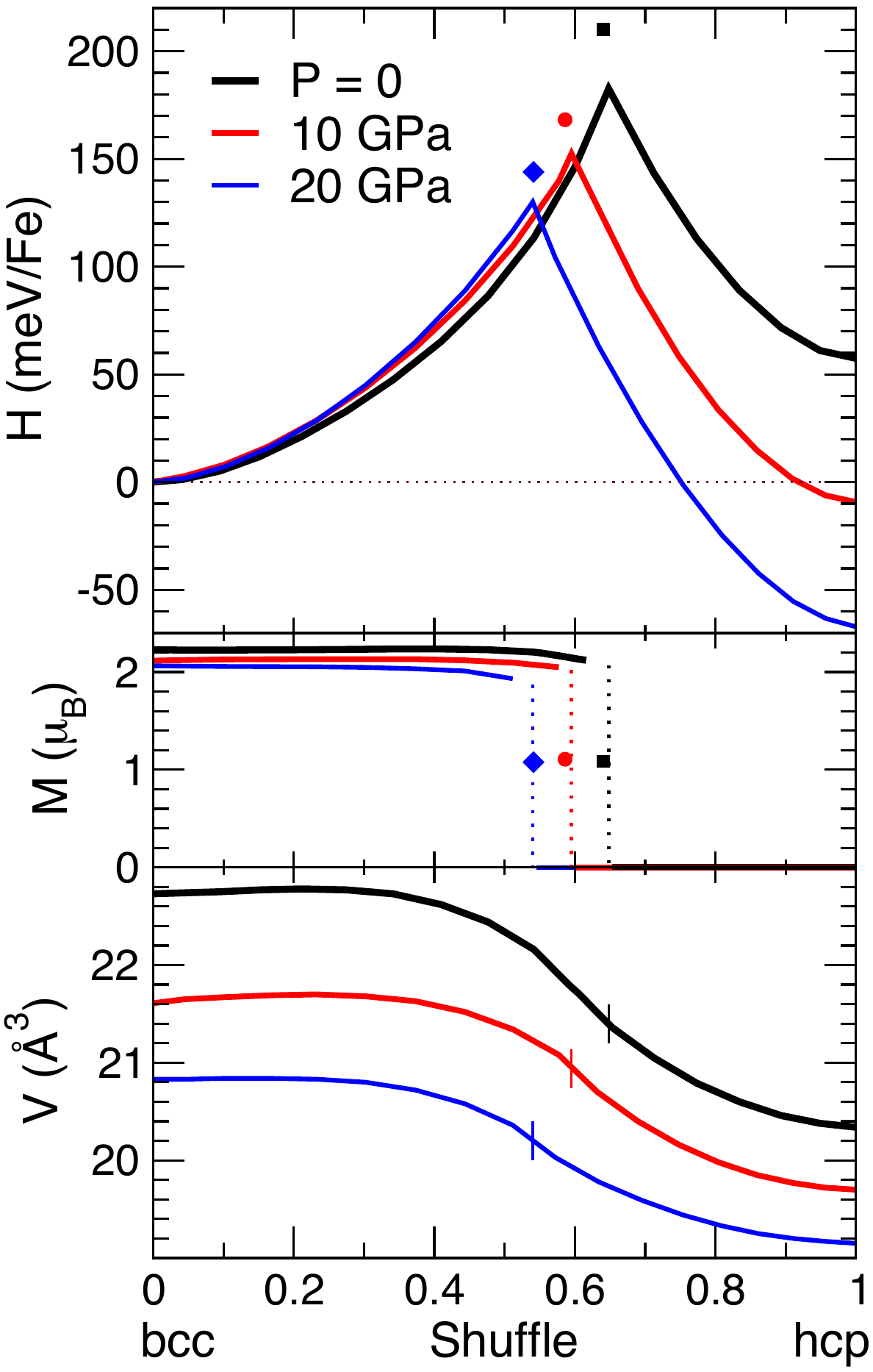} 
\vspace{-2mm}
\caption{\label{Fig2}(color online) 
For 2-atom cell, enthalpy $H$ [meV per atom] relative to bcc Fe, magnetization M [$\mu_B$ per atom], and volume V [{\AA}$^3$ per unit cell]  vs. shuffle from SSNEB at 3 hydrostatic pressures: $P$=0 (black), 10 (red), and 20 GPa (blue). Higher-enthalpy states with intermediate $M$ are shown as filled shapes. Dotted vertical lines and dashes are at discontinuities in $M$. 
}
\end{figure}
\begin{figure}[t]
\includegraphics[width=60mm]{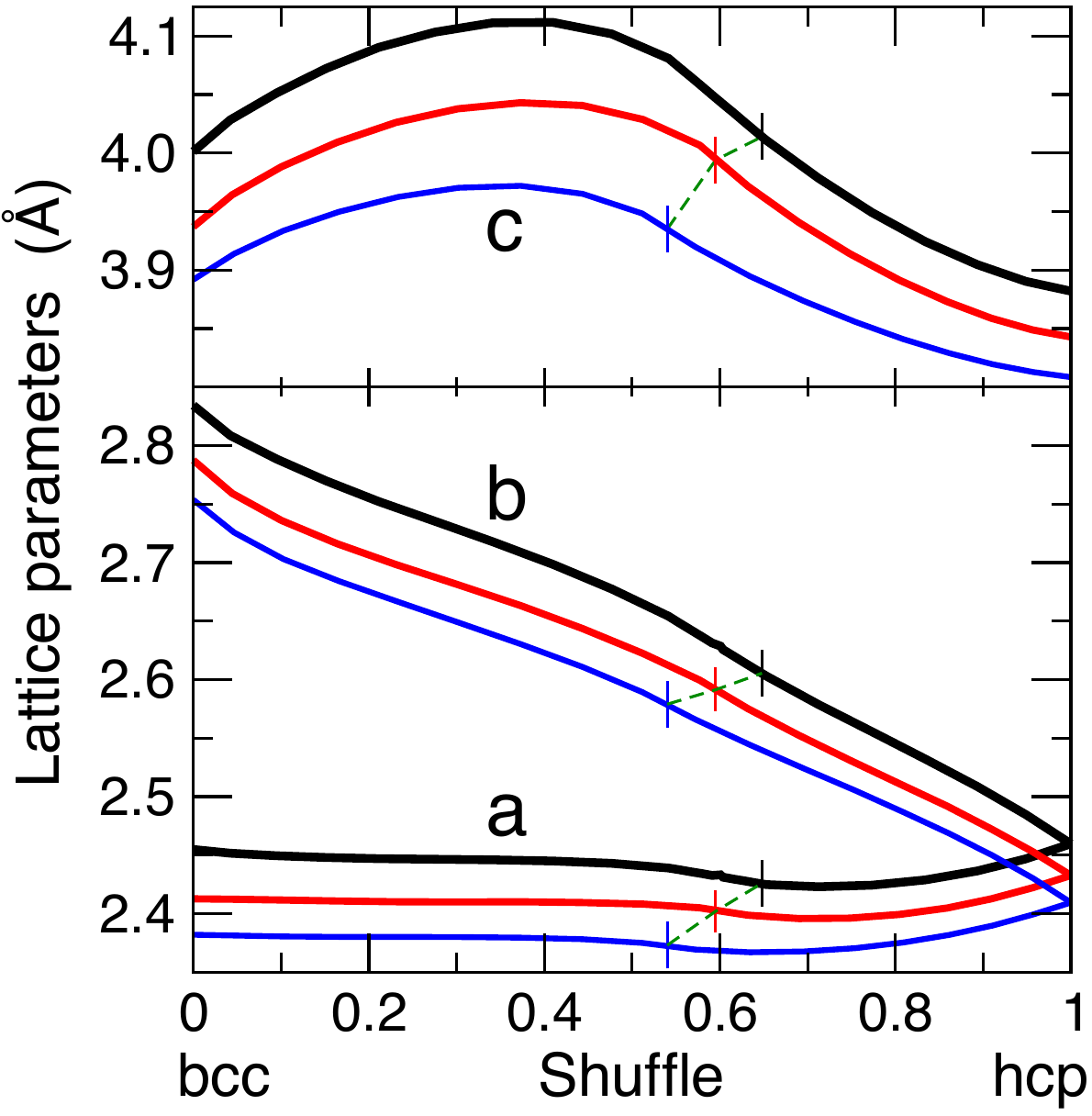} 
\vspace{-2mm}
\caption{\label{Fig3}(color online) 
Lattice parameters $a$, $b=2 a \, \sin (\vartheta /2)⁡$, and $c$ (\AA) versus shuffle for 3 values of external pressure (GPa): $P$=0 (black), 10  (red), and 20 (blue). Trajectory of the TS is given by dashed line. 
}
\end{figure}

{\par }
The 3 independent cell DoF for an orthorhombic unit cell are the lengths of the 3 mutually orthogonal vectors $\bf{a}_1 + \bf{a}_2$, $\bf{a}_2 - \bf{a}_1$, and $\bf{c}$ 
or 3 independent functions of them (e.g., cell volume $V$, shear angle $\vartheta$, and $c/a$). 
We define diagonal vector $\bf{b} = \bf{a}_2 - \bf{a}_1$, which is orthogonal to $\bf{c}$ and 
shuffle $\bf{s}$ (directed along the other diagonal $\bf{a}_1 + \bf{a}_2$), see Fig.~\ref{Fig1}. 
We emphasize that the complete description of this bcc-hcp transformation requires consideration of all 4 independent atomic and cell DoF.

{\par }Previously, the bcc-hcp transformation path was usually described by oversimplified models  \cite{Science238n4828y1987,JChemPhys128n10p104703y2008}.
In particular, the shuffle-shear model \cite{Science238n4828y1987} was
a drag method that ignored critical changes in magnetization and volume and hence 
improperly uncoupled atomic and cell DoF. 
A drag method improperly allows for discontinuity in some of the uncontrolled DoF; hence, the resulting transformation path can ``tunnel'' under the TS (due to discontinuous DoF) and bypass it, missing and hence underestimating the true enthalpy barrier \cite{GSSNEB}. 
Use of a drag method based on the rapid-nuclear-motion (RNM) approximation leads to a discontinuity in atomic shuffle, 
giving a very low bcc-hcp barrier in the form of a cusp, with equal enthalpies of the bcc and hcp phases at the calculated pressure of 13.1 GPa  \cite{JChemPhys128n10p104703y2008}.
We emphasize that the shuffle-shear model \cite{Science238n4828y1987}, which is a drag method, controls of only 2 DoF (shuffle and shear), leaving an uncontrolled unit cell volume beyond its consideration \cite{PRB79n13p134113}.

{\par }
The iron bcc-hcp transformation path was calculated directly using density functional theory (DFT) for the shuffle-shear model in a 2-atom cell (Fig.~\ref{Fig1}), 
with full volume relaxation for each pair of shuffle-shear values \cite{PRB79n13p134113}, finding a MEP with a cusp due to the magneto-volume collapse. 
In addition to a discontinuity in volume,  the reported MEP cusp was rounded due to the use of an approximate symmetry-adapted polynomial to  
analytically determine the entire potential energy surface.  
More recently, possible concerted transformations were investigated (including 3 shuffling mechanisms within the RNM at constant shear and fixed 71.5~bohr$^3$/atom volume), finding a TS in the form of a cusp \cite{PRB87n2p024103y2013}, but missing the observed magneto-volume collapse --  disappearance of magnetization with a concomitant $11$\% drop in volume.

{\par }
Classical inter-atomic potentials typically ignore dependence on the magnetic state, giving an irrelevant path for magneto-structural transitions. 
Thus, the embedded atom method (EAM) provided an overestimated transition pressure of 31--33 GPa for the uniform \cite{JPhysCM22n43p435404y2010} and 14 GPa for the uniaxial compression \cite{JPhysCM21n49p495702y2009}.

{\par }
While the  pressure-induced bcc-hcp transformation in iron has been extensively studied, its complete theoretical description accounting for all relevant DoF is still needed, and we provide it below. 
Recall that the traditional SSNEB method \cite{GSSNEB} can search the entire phase space, but requires a continuous and differentiable PES with the TS in the form of a saddle 
(e.g., as in Li \cite{PNAS102p6738y2005}),  which can be addressed by the available codes \cite{C2NEBsoft} with one \cite{C1NEB} or (more stable) two \cite{C2NEB} climbing images, 
while for a magneto-structural transformation the TS is a cusp on the PES with a discontinuity of atomic forces and stress components, where a climbing image will not stop.  
We extend the unrestricted SSNEB formalism (Appendix \ref{sCompMethod}) and apply it to the bcc-hcp transformation in iron under pressure. 
We compare our results to experiment and contrast them with the previous theoretical discussions.

\section{\label{Results}Results}
\begin{figure}[t]
\includegraphics[height=73mm]{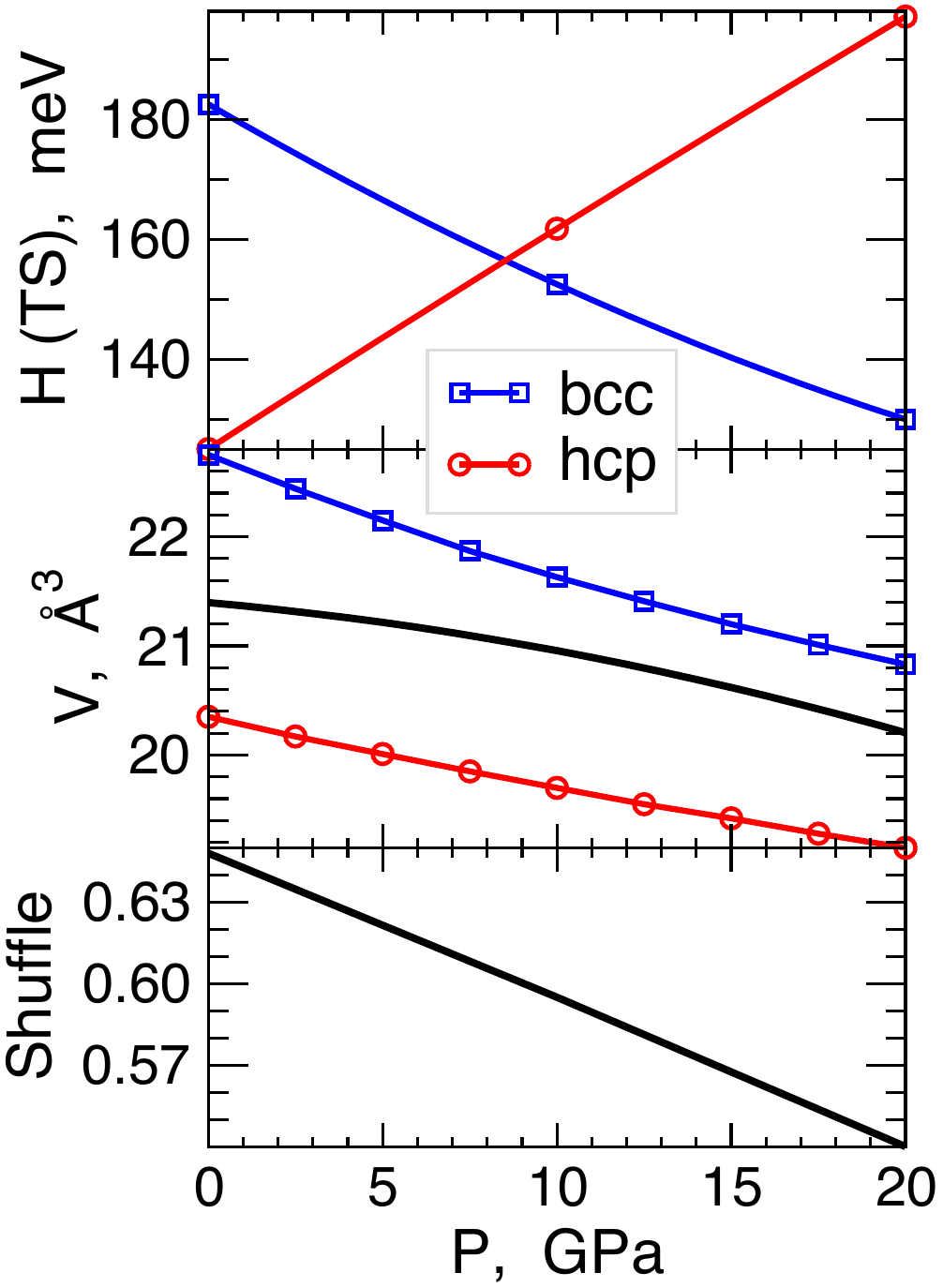} 
\vspace{-2mm}
\caption{\label{Fig5}(color online) 
Versus external pressure $P$, enthalpy [meV/Fe] of the TS (156 meV/Fe at $P_0 =$8.42$\,$GPa) relative to bcc (blue) and hcp (red); unit cell volume of the TS (black), bcc (blue), and hcp (red) structures; and shuffle  at the TS.   
}
\end{figure}

\begin{figure}[t]
\includegraphics[height=73mm]{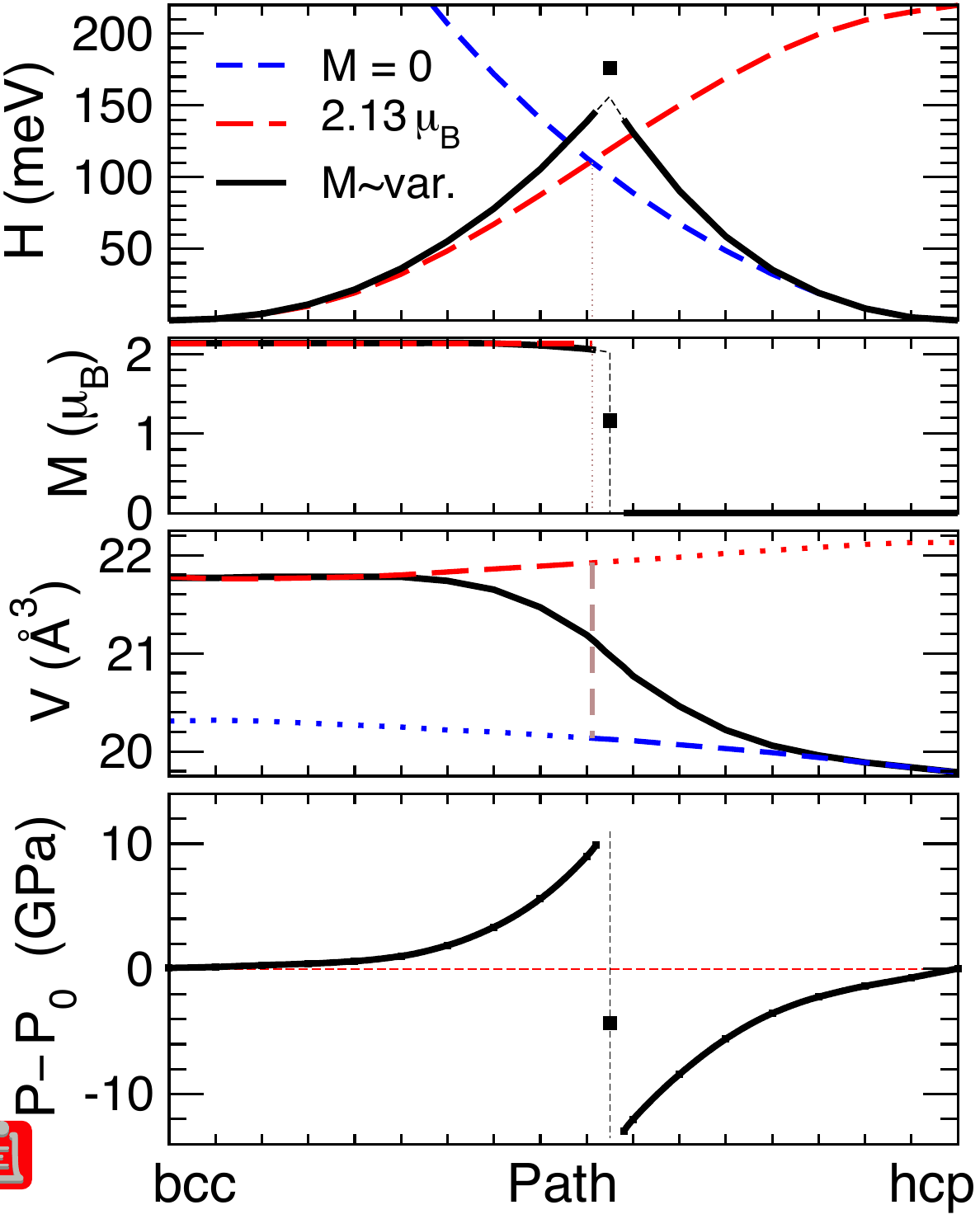} 
\vspace{-2mm}
\caption{\label{Fig7}(color online) 
For 2-atom cell, SSNEB enthalpy $H$ [meV/atom],  magnetization $M$ [$\mu_B$/atom], volume $V$ [{\AA}$^3$ per cell], and internal pressure $P$ [GPa], versus  MEP.  SSNEB results for variable $M$ (analytic continuation of $H$ near cusp)  are compared to those at fixed $M$, i.e., $M=0$ (blue short-dash line) and $2.13 \, \mu_B$/Fe (red long-dash line).  Two fixed-M  solutions never cross on the PES, so there is a jump between them (vertical brown dashed line, as seen in $V$ vs. path)  to avoid higher-enthalpy solutions (blue and red dotted lines).  Higher-enthalpy intermediate state (black squares) with intermediate $M$ arises from improper VASP convergence.}
\end{figure}

{\par} Using our modified SSNEB method (see Appendix \ref{sCompMethod}) with properly coupled cell and atomic DoF, we perform calculations at several values of the applied hydrostatic pressure $P$, including 0, $8.4$, 10, and 20 GPa. 
Because an equilibrium coexistence is rarely observed in experiment, here we provide the barrier versus pressure (Figs.~\ref{Fig2}, \ref{Fig3}, and \ref{Fig5}), emphasizing generality of our  quantitative description.
The result at the equilibrium coexistence $P_0$=8.4$\,$GPa is given in Fig.~3 in \cite{PRB_ironP} (see also our Fig.~\ref{Fig7}).


\subsection{Atomic and cell degrees of freedom}
{\par}The 2-atom cell has 6 atomic DoF (3 of which are independent) and 6 cell DoF (3 lengths of the lattice translation vectors and 3 angles between them). We find that out of the 3 independent atomic DoF, only 1 (shuffle) is interesting, while the other two can be chosen in such a way that they remain constant during the whole transformation. Among the 6 cell DoF, values of 2 out of 3 angles remain constant ($90^\circ$). The cell can be chosen with a mirror symmetry, so that 2 lengths of the translation vectors are equal (Fig.~\ref{Fig1}). 
We are left with 1 changing atomic DoF (shuffle) and 3 cell DoF.  The direct coordinates of atoms are (0, 0, 0) and ($\frac{1}{2} [1+\frac{s}{3}$]; $\frac{1}{2} [1+\frac{s}{3}$]; $\frac{1}{2}$), 
so that the second ones change from ($1/2$, $1/2$, $1/2$) in bcc to ($2/3$, $2/3$, $1/2$) in hcp (Fig.~\ref{Fig1}). 

{\par} 
We find that the atomic shuffle $s$ is a continuous monotonic function of the SSNEB path variable, hence all the other
path-dependent variables can be expressed in terms of the shuffle.
All 3 independent cell DoF are plotted in terms of the shuffle in Fig.~\ref{Fig3}, where we introduced $b=2 a \, \sin⁡(\vartheta /2)$ as the length of $\bf{b}=\bf{a}_2-\bf{a}_1$ (Fig.~\ref{Fig1}). 
Figures~\ref{Fig2} and \ref{Fig3} show that although there is a discontinuity in magnetization at the TS, 
all the atomic and cell DoF are continuous functions versus the MEP (and the shuffle). 
This is in contrast to drag methods that have unphysical discontinuities in some of the uncontrolled DoF. 
At P$_0 = 8.4\,$GPa, the TS is characterized by $a=2.41\,${\AA}, $b=2.59\,${\AA}, $c=3.99\,${\AA}, and $s=0.6$, with $V=21.0\,\mbox{\AA}^3$, $\vartheta = 65.1^\circ$, and $c/a = 1.6594$.  

{\par}Although the SSNEB path depends on the Jacobian $J$, coupling atomic and cell degrees of freedom \cite{GSSNEB},  
the TS is invariant for reasonable finite values of $J$ (typically between 1 and 5). 
At any $J$, the SSNEB path goes through the TS and the terminal states.
With $J\approx 3$, we find an approximately linear dependence between the atomic shuffle ${s}$ and cell translation ${b}$  (Fig.~\ref{Fig3}). 
Dependence of the shuffle at the TS on the hydrostatic external pressure $P$ is nearly linear (Fig.~\ref{Fig5} bottom).
At the equilibrium coexistence pressure P$_0$=8.42 GPa, where bcc and hcp phases have equal enthalpies \cite{PRB_ironP}, the calculated enthalpy of the TS is 156 meV/atom above the terminal states.

{\par} We emphasize that this value is the same 
from the intersection of lines connecting the SSNEB results at $0$, 10, and 20 GPa (Fig.~\ref{Fig5}), 
from our SSNEB calculation at $P_0$ (Fig.~\ref{Fig7}), and
from direct DFT calculations of the electronic structure (Fig.~\ref{Fig6}) and enthalpy of the TS atomic configuration at $P_0$ (Fig.~\ref{Fig8}).

\subsection{Electronic structure and magnetization}
\begin{figure}[t]
\includegraphics[width=75mm]{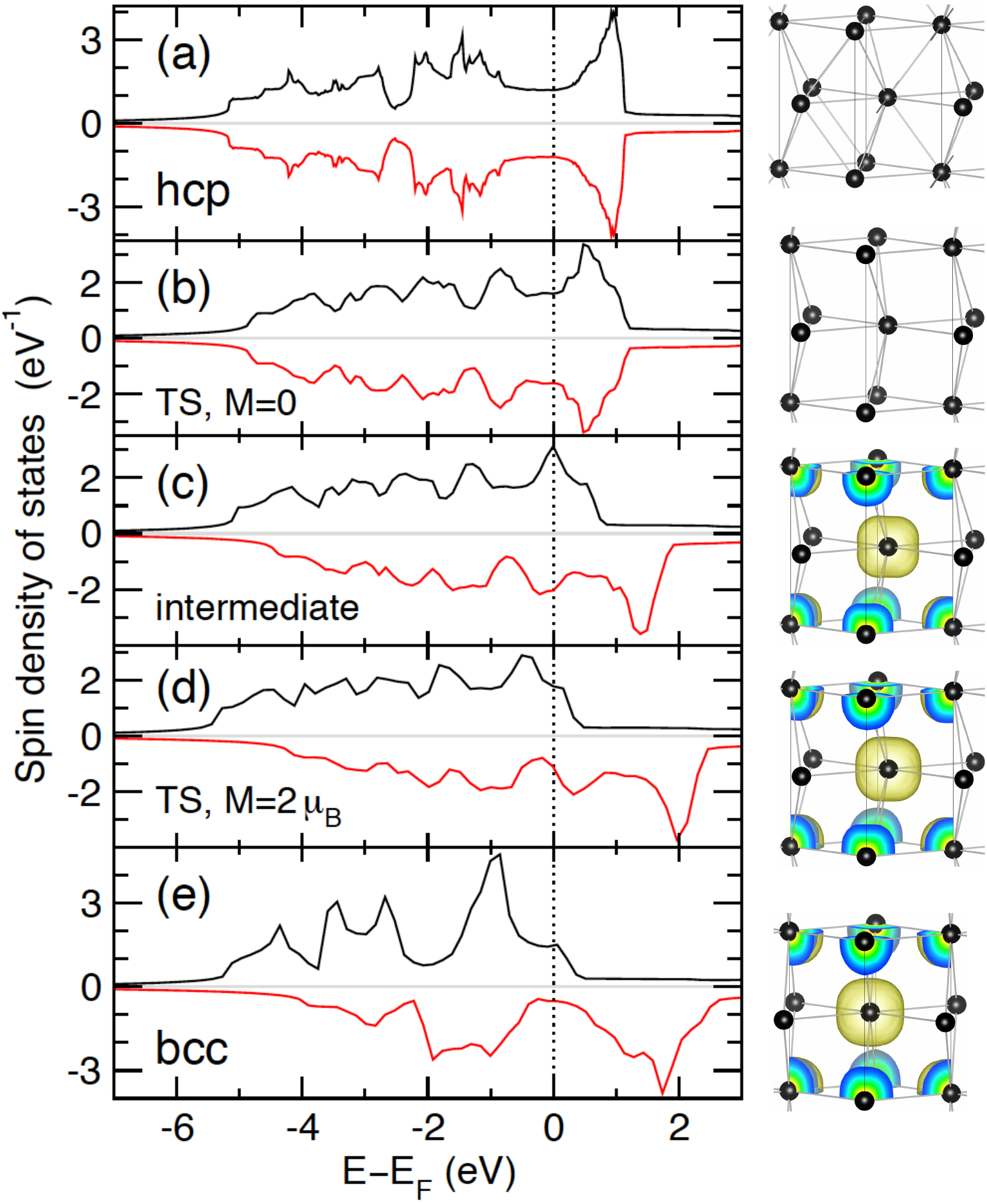} 
\vspace{-2mm}
\caption{\label{Fig6}(color online) 
Total spin DOS [states per atom per eV] for the bcc-hcp transformation at $P_0=8.4\,$GPa. 
Majority (minority) spin states are plotted on positive (negative) axes.
Endpoint are (a) non-magnetic hcp and (e) magnetic ($M \! = \! 2.12\, \mu_B$/Fe) bcc. 
Transition states (TS) are (b) non-magnetic ($M\!=\!0$) and (d) magnetic ($M\!=\!2\, \mu_B$/Fe), 
 having same atomic structure and enthalpy (the cusp in Fig.~\ref{Fig7}). 
(c)  Higher-enthalpy intermediate state ($M \! = \! 1.1\, \mu_B$/Fe) with the same atomic structure as the TS. 
States (b, c, d) are enthalpy extrema, marked (red squares) in Fig.~\ref{Fig8}. 
The right panels show inter\-atomic bonds up to $2.5\,${\AA} and $0.02\,e^-/\mbox{\AA}^3$ iso-surfaces of the spin density for the magnetic structures. 
}
\end{figure}

\begin{figure}[t]
\includegraphics[width=70mm]{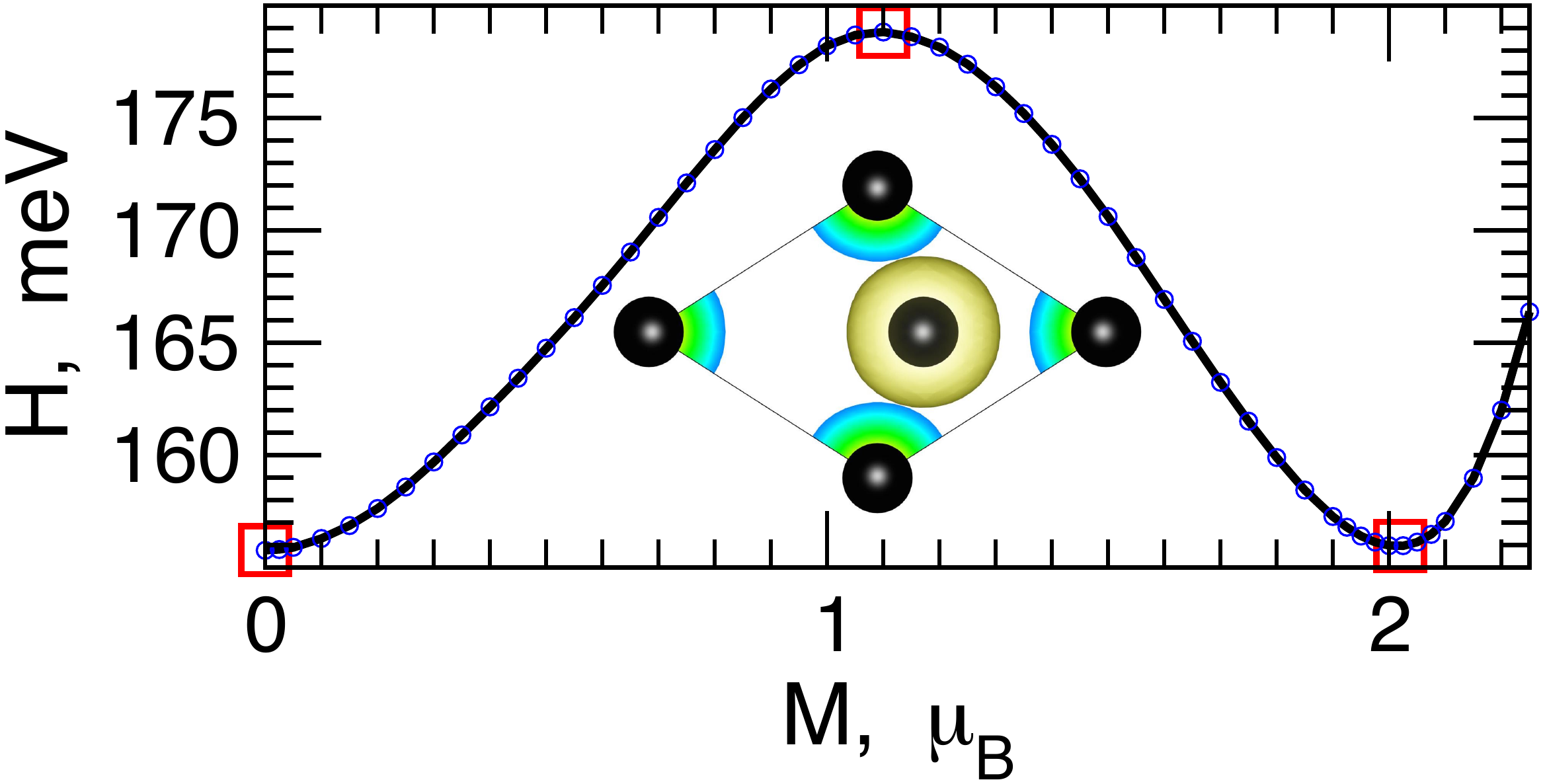} 
\vspace{-2mm}
\caption{\label{Fig8}(color online) Enthalpy (meV/atom) vs. magnetization $M$ ($\mu_B$/atom) relative to the equal enthalpies of bcc and hcp at $P_0=8.4~$GPa for the transition and intermediate states with the same atomic structure. The extrema are marked by red squares. The inset shows atomic structure with $0.2\,e^-/\mbox{\AA}$ iso-surfaces of the total electron density for the magnetic TS. }
\end{figure}

{\par } A magneto-structural transformation is characterized by a duality of the TS, with the same values of all the cell and atomic coordinates and the same enthalpy, but different magnetization (Fig.~\ref{Fig2}) and different electronic structure and density of states (DOS, Fig.~\ref{Fig6}).
 Each intermediate (higher enthalpy) state with a fractional magnetization (filled shapes in Fig.~\ref{Fig2}) can be regarded as an electronically excited state (Fig.~\ref{Fig8}).   
The DOS of the intermediate state is compared with DOS of two TS and two endpoints in Fig.~\ref{Fig6}. The bcc and hcp end points and both TS have local minimum of DOS at the Fermi level. 
Bands continuously change between the end point and the corresponding TS, but the Fermi surface remains at the local minimum of the total DOS. Although the bands are not rigid, and the rigid-band model \cite{bookRigidBand1958} is not precise here, the difference between two TS is roughly a band shift, with a local maximum in majority and minority spins passing the Fermi surface.  The higher-enthalpy intermediate state has that band at the Fermi surface. 

{\par}  Magnetization is not a conserved DoF and is allowed to have discontinuity at the TS. 
DFT is expected to converge to a local enthalpy minimum for each atomic structure, including the TS. 
Figure~\ref{Fig8} shows two such minima (non-magnetic at $M=0$ and magnetic at $M \! \approx \! 2\,\mu_B$/Fe) with the same enthalpy but different values of magnetization;  they are separated by the higher-enthalpy intermediate states, which can be regarded as electronically excited states with different values of $M$.  
Recall that both TS and intermediate states have the same atomic structure (identical atomic positions and lattice vectors), but different electronic structure.

{\par} Notably, the directly calculated $H$ vs. $M$ (Fig.~\ref{Fig8}) indicates that the states with intermediate magnetization (obtained by DFT with fixed $M$) have unstable electronic structure, but reflect an accommodation for the allowed discontinuity in $M$ at the TS.  
Two electronically stable states at the $H$ minima in Fig.~\ref{Fig8} provide the enthalpy barrier, while all the others (including the maximal $H$) do not, because they are electronically unstable.
{DFT is expected to find a local enthalpy minimum for any structure, including the TS; however, we find that the widely used DFT code VASP \cite{VASP1,VASP2} converges the electronic structure to a local enthalpy extremum, which is \emph{maximum} near the TS and minimum elsewhere.}
Careful application of the SSNEB approaching the TS (cusp region) avoids those ``intermediate'' electronically excited states (filled shapes in Fig.~\ref{Fig2}) due to the additional stop condition, discussed in Sec.~\ref{recursiveNEB} and \ref{NEB2TS}.

\subsection{Comparison to drag methods}
{\par } Most previous attempts to model the bcc-hcp transformation in iron ultimately  implemented a drag method, with discontinuity in one or more DoF. 
The RNM approximation \cite{JChemPhys128n10p104703y2008} results in a discontinuity in atomic shuffle. 
The shuffle-shear model \cite{Science238n4828y1987} allows discontinuity of volume and lattice parameters  \cite{PRB79n13p134113}.
By selecting too small (zero) or too large (infinite) Jacobian $J$ within the  SSNEB \cite{GSSNEB}, we can ignore either atomic (as in RNM) or cell DoF and reproduce the results of those drag methods with discontinuity in the ignored DoF, included in SSNEB with a negligible weight $J \to 0$ or $1/J \ll 1$. 

{\par }With a discontinuity in any DoF, the physical system is dragged under an energy barrier; the true barrier is bypassed, and the calculated fictitious barrier is typically lower than the actual physical one. 
On the other hand, fixing or limiting selected DoF imposes restrictions on possible transition paths, constraining the search space, 
so that the minimum energy path can be missed: in this case the calculated barrier can be higher than the TS on the MEP. 

{\par } To avoid dealing with a cusp, it is tempting to consider two solutions with fixed magnetization (Fig.~\ref{Fig7}). 
Although two \emph{projected} energy curves versus path have \emph{apparent} intersection in Fig.~\ref{Fig7},
this is not a TS, because two sets of DoF do not coincide anywhere.  In other words, we find that fixed-magnetization solutions with $M=0$ and $2.13 \, \mu_B$/Fe do not intersect in a multi-dimensional space parametrized by all DoF. 
In particular, unit cell volume of the magnetic solution is larger than that of the non-magnetic one at every point in Fig.~\ref{Fig7}. 
The apparent intersection of the energy vs. path projected curves is a fictitious barrier (with an improper discontinuity in volume and lattice constants), which is lower than the real one.  At fixed magnetization, energy is a smooth function of DoF; and pressure and atomic forces can be converged to zero everywhere along such paths.


\subsection{Pressure discontinuity at the TS}
{\par }In contrast, pressure can be discontinuous at the TS.  
Two states with two different magnetization values are forced to have the same atomic and cell coordinates. Both are strained, and those strains have equal amplitudes and opposite directions in the dual TS. The TS relaxation to zero pressure would create discontinuity in one or more DoF (this was previously done in drag methods). In particular, continuity of volume creates discontinuity of pressure, and vice versus. 

{\par }Interestingly, pressure discontinuity of 25 GPa (from $-13$ to 12 GPa in Fig.~\ref{Fig7}) at the TS  has the same order of magnitude as the spread of experimental pressures for coexisting bcc and hcp phases (from 0 to 25 GPa). This is not a coincidence. Additional broadening of the measured pressures is caused by the martensitic stress.

{\par } Indeed, the bcc-hcp ($\alpha$-$\varepsilon$) transition in iron is martensitic, and the hysteresis loop can be  characterized by four pressure values: $P_{start}^{(\alpha \to \varepsilon)}$ and $P_{end}^{(\alpha \to \varepsilon)}$ for direct bcc$\to$hcp; 
$P_{start}^{(\varepsilon \to \alpha)}$ and $P_{end}^{(\varepsilon \to \alpha)}$ for the reverse hcp$\to$bcc transform \cite{PRB_ironP}.
In one experiment \cite{Science238n4828y1987}, the hcp phase appears at $P_{start}^{(\alpha \to \varepsilon)}=10.8 \pm 0.5 \,$GPa and  bcc  disappears above $P_{end}^{(\alpha \to \varepsilon)}=21 \,$GPa upon loading; upon unloading, bcc reappears at $P_{start}^{(\varepsilon \to \alpha)}=15.8 \pm 0.5 \, $GPa and  hcp disappears below $P_{end}^{(\varepsilon \to \alpha)}=3 \, $GPa .  
The  observed bcc-to-hcp onset $P_{start}^{(\alpha \to \varepsilon)}$ has been reported from $8.6$ to $15$~GPa, with the highest experimental $P_{end}^{(\varepsilon \to \alpha)}=8.5 \pm 0.6 \, $GPa \cite{PRB_ironP}.
Our calculated hydrostatic equilibrium coexistence pressure of 8.4 GPa agrees well with a range of experimental measurements \cite{PRB_ironP}.
The 25~GPa pressure discontinuity at the TS agrees with the experimental $\approx 25 \,$GPa spread of the observed bcc--hcp coexistence (Table~1 in \cite{PRB_ironP}).

\section{\label{Summary}Summary}
{\par } In summary, we extended the SSNEB formalism to address magneto-volume collapse in magnetic materials under stress or temperature, caused by loss of magnetization along the transformation pathway, where magnetization is a non-conserved order parameter, unlike the continuous atomic and cell degrees of freedom.  We applied it to the pressure-induced bcc-hcp magneto-structural transformation in iron, which exhibits a well-known 11\% volume decrease concomitant with the loss of magnetization. We contrasted DFT-based equilibrium coexistence pressure (typically found by Maxwell construction) and ``apparent coexistence'' pressures typically measured, which are highly affected by internal stresses in the samples.
For iron under pressure, we found the transition state in the form of a cusp on the potential energy surface, although all the  atomic and cell degrees of freedom are continuous, as physically required.
We explained the difference between the generalized SSNEB and all previously used drag and restricted parametrization methods. 

{\par} Our calculated values of the bcc-hcp equilibrium coexistence pressure, energy barrier, and pressure discontinuity at the transition state all agree with the experimental data. 
The new formalism is suitable for studying numerous magnetic systems, especially those involving magneto-structural transformations, 
such as transducers, magnetic switches, and competing chemical and magnetic structures.
Our extended formalism works for barriers in the forms of both saddle points and cusps.

\acknowledgments
We thank  Anatoly Belonoshko, Igor Abrikosov, and Iver Anderson for useful discussions. 
This work was supported by the U.S. Department of Energy, Office of Basic Energy Sciences, Division of Materials Science and Engineering. The research was performed at the Ames Laboratory, which is operated for the U.S. DOE by Iowa State University under contract DE-AC02-07CH11358.

\appendix

\section{\label{sCompMethod} Computational Methods}
{\par }
	We combine density functional theory (DFT, \ref{DFT}) with the extended nudged elastic band methods (\ref{recursiveNEB} and \ref{NEB2TS}).  

\subsection{\label{DFT}DFT}
{\par} 
 The Vienna {\em ab initio} simulation package VASP \cite{VASP1,VASP2}
is used to calculate electronic energy, pressure, and atomic forces
for instantaneous atomic configurations.  We use the projector augmented wave (PAW) \cite{PRB50p17953} technique, the generalized gradient approximation (GGA \cite{GGA}), 
 a plane-wave energy cutoff of 700~eV, and a
converged $k$-point mesh \cite{MonkhorstPack1976} including $\Gamma$-point for the Brillouin zone
integration with at least 75 points per \AA$^{-1}$  (e.g., $17$ $k$-points for $b=4.4\,$\AA). 
The modified Broyden method \cite{PRB38p12807y1988} is used for self-consistency.
Atomic and cell relaxations (including those within the NEB method) are performed by selective dynamics
using a conjugate-gradient algorithm with a Gaussian smearing with $\sigma = 0.05\,$eV. 
Structural energy differences are obtained using the tetrahedron method with Bl\"ochl corrections.

{\par } For any atomic structure, DFT should find a local energy minimum in terms of the electronic density.  
VASP does it for most structures, except those near the TS.  
At the TS  it converges to a local enthalpy maximum (Fig.~\ref{Fig8}) --- 
an intermediate state with a higher enthalpy and a fractional magnetization (filled symbols in Figs.~\ref{Fig2} and \ref{Fig7}).  
Convergence to the nearest extremum (which is expected to be a minimum) provides an unexpected result (maximum) near the TS.
At present, we have no explanation for this feature.

\subsection{\label{recursiveNEB}NEB with recursively added images near TS}
{\par }  We recursively add images near the TS within the SSNEB method \cite{GSSNEB}.
First, a traditional SSNEB \cite{GSSNEB} with equidistant images is used. 
The input configuration is prepared by adjusting initial unit cell volume of each image in such a way, that all images between the magnetic bcc end point and the expected TS are ferromagnetic (FM), while all images on the other side (TS to hcp) have zero magnetization. 
After convergence of this SSNEB calculation, we take two closest to the TS neighbor images with different magnetic states, 
and use them as new terminal points for the next SSNEB calculation with a sufficiently large number of images. 
Adding images between those closest to the TS is repeated until the cusp is well-sampled by a dense grid.
We stop these iterations when any image added between the two TS neighbors (one is ferromagnetic, the other is non-magnetic)
gets an intermediate magnetization $0.1<M<1.8 \, \mu_B$/atom.

\subsection{\label{NEB2TS}NEB with two images approaching the TS}
{\par}
The traditional SSNEB method \cite{GSSNEB}  
expects a TS to be a saddle point of the smooth potential energy surface. In the single climbing image algorithm \cite{C1NEB}, the highest enthalpy image is driven up to the saddle point. This image does not feel the spring forces along the band. It tries to maximize its energy along the NEB, and minimize it in all other directions. This algorithm requires continuity of forces and pressures along the path, and expects zero forces at the saddle point; it is not suitable for PES with a barrier in the form of a cusp, where discontinuity of pressure and atomic forces is at issue.

{\par}
	In contrast, magnetic and magneto-structural transformations can be addressed by a modified SSNEB, where the TS is approached by two images with different magnetization (Fig.~\ref{Fig2}), 
converging towards the same values of all atomic and cell DoF and the same enthalpy. 
Ferromagnetic and non-magnetic states form two smooth energy surfaces; the minimal energy surface forms a cusp at their intersection; 
 components of the atomic forces and stress tensor have discontinuities at that cusp (Fig.~\ref{Fig7}). 

{\par}
In our algorithm, except for the two terminal images and two images near the TS, all the other images are nudged. 
	At each step, a trial image is placed between the two images with true tangent force projections having opposite directions (those two images bracket the TS from two sides). 
Initial coordinates of the trial image are the weighted average of coordinates of those two images: 
$I_{Trial} = m I_1 + (1-m) I_2$, 
where the weight $0<m<1$ can have any value between zero and one (we chose $m=1/2$).  
The trial image is relaxed to the point with zero (or small, below the specified cutoff) perpendicular forces. 
The true tangent force projection is calculated for the trial image and its direction is compared with those for its two neighbors. 
The image with tangent force in the same direction is replaced by the trial image, so that the tangent forces of two climbing images have opposite directions again. 
Calculation stops if
magnetization of any image has an unacceptable intermediate value, or 
 the ``distance'' between two images is below a certain cutoff. 
The normalized ``distance'' between images is the square root of the sum of weighted squares of the differences between components of atomic coordinates, lattice translation vectors, and two enthalpies.  
At each step, only one trial image is computed and only one of the two TS neighbors is moved.

{\par}Our algorithm works for barriers in the forms of both saddle points and cusps. 
It can safely replace the climbing-image algorithm C1-NEB \cite{C1NEB} in all cases, where C1-NEB is applicable, and can be generalized for complex energy landscapes \cite{C2NEB}.
However, the SSNEB convergence malfunctions if any image gets an excited electronic state with intermediate magnetization (Fig. \ref{Fig6}). 
Fortunately, such cases can be easily detected, and we monitor them by an additional check: the calculation stops if any image has an unacceptable intermediate magnetization, $0.1<M<1.8 \, \mu_B$/atom. As a result, we get two images with very similar values of enthalpy and all atomic and cell DoF, and an intermediate state with a higher enthalpy between them. A similar result was obtained from a dense grid of images near the TS in section \ref{recursiveNEB}.

\bibliography{Fe}
\end{document}